# One-Dimensional Energy Dispersion of Single-Walled Carbon Nanotubes by Resonant Electron Scattering

Min Ouyang, Jin-Lin Huang, and Charles M. Lieber[*]
*Harvard University, 12 Oxford Street, Cambridge, MA 02138*

**Abstract**

We characterized the energy band dispersion near the Fermi level in single-walled carbon nanotubes using low-temperature scanning tunneling microscopy. Analysis of energy dependent standing wave oscillations, which result from quantum interference of electrons resonantly scattered by defects, yield a linear energy dispersion near $E_F$, and indicate the importance of parity in scattering for armchair single-walled carbon nanotubes. Additionally, these data provide values of the tight-binding overlap integral and Fermi wavevector in good agreement with previous work, but indicate that the electron coherence length is substantially shortened.

    The remarkable electronic properties of single-walled carbon nanotubes (SWNTs), which are due in large part to their unusual band structures [1-7], have aroused considerable excitement in fundamental and applied research [1-3]. In the case of isolated armchair (n,n) SWNTs, the $\pi$-bonding ($\pi$) and $\pi$-antibonding ($\pi^*$) energy bands are predicted to cross at the Fermi level ($E_F$) in an unique linear fashion, contrasting the parabolic dependence expected from a conventional free-electron picture. However, experimental determination of this important characteristic of the band structure has been lacking. Use of conventional momentum analysis methods, which average over substantial area, is difficult since SWNT samples consist of a wide range of structures each with different energy dispersions [5,6]. Scanning tunneling microscopy (STM), which can be used to interrogate individual nanotubes and has been used previously to determine the energy dispersions of 2D surface states on metals [8,9] and semiconductors [10,11], could thus address this critical point.

    Here we report STM studies that elucidate the 1D energy dispersion of SWNTs by spatially resolving energy dependent standing wave oscillations near an isolated defect. Our STM measurements were performed in a home-built, ultrahigh vacuum STM at ~5 K. The SWNT samples were prepared by spin-coating 1,2-dichloroethane suspensions of SWNTs onto Au(111) surfaces, and current-voltage (*I*-V) and conductance-voltage (*(dI/dV)*-V) spectra were recorded simultaneously using published methods [3,5,7]. An image of a metallic SWNT with a defect located ca. in the middle of the image (Fig. 1a) exhibits a strong modulation in the local density of states (LDOS) with a period larger than that of the graphene lattice. Tunneling spectroscopy and imaging measurements made ca. 8 nm away from either side of the defect, where these modulations have decayed, exhibit van Hove singularities (VHS) (Fig. 1a) and an atomic lattice characteristic of a (13,13) armchair SWNT [3,12].

    Central to the focus of our work is the use of STM to map in detail the modulations in the LDOS as a function of energy and spatial position with respect to the defect in Fig. 1a [13]. A series of experimental *(dI/dV)/(I/V)* curves (Fig. 1b), which are proportional to the LDOS, recorded as a function of both energy and distance from the defect show nine peaks whose amplitudes vary with position moving away from the defect. These features in the LDOS occur

within the first VHS, and moreover, are not present in spectra recorded far away from the defect. The most striking aspect of these peaks is the spatial oscillation of their amplitudes along the SWNT axis. Specifically, the amplitude of each of these peaks varies periodically with position and decays on a nanometer length scale. The oscillatory behavior of these features in the LDOS can be seen clearly by plotting the amplitude vs. position for specific energies (Fig. 2). These data also show that the oscillation wavelength decreases as the magnitude of the energy increases with respect to $E_F$.

Similar spatial oscillations in the LDOS have been reported on 2D surfaces and were attributed to the interference of non-resonantly scattering surface state electrons from surface imperfections [8-11]. For 1D metallic SWNTs with a defect, previous theoretical studies [14] and transport measurements [15] have shown that resonant electron scattering can occur when the incident electron energy matches (is resonant with) a defect related quasi-bound state. Our experimental observations of oscillations occurring only at specific energies are consistent with these studies and strongly suggest that quantum interference between forward propagating and resonantly backscattered electrons can explain the observed oscillations [16].

In this vein, we model the observed oscillations in the armchair SWNT using a 1D plane-wave $e^{ikx}$, where $k$ is wavevector and x is position, with an energy equal to the defect state and incident on the defect at x = 0. The incoming wave is reflected by resonant backscattering, with a reflectivity $|R|^2$ ($R = |R|e^{-i\delta}$, where $\delta$ is the phase shift), and interferes with the incident plane-wave (Fig. 2a). The resulting standing wave-function can be written as $\psi(k,x) = e^{ikx} + |R|e^{-i(kx + \delta)}$, leading to spatial oscillations in the LDOS, $\rho(k,x)$, at the energies of the defect states:

$$\rho(k,x) = |\psi(k,x)|^2 = 1+|R|^2+2|R|\cos(2kx+\delta) \qquad (1)$$

In addition, processes such as electron-electron (*e-e*) and electron-phonon interactions can dephase electron wavefunction and lead to a damping of the interference [17]. We include dephasing with the exponential damping term $e^{-x/l_\varphi}$, where $l_\varphi$ is a phenomenological coherence length, which leads to a modified expression for Eq. 1:

$$\rho(k,x)= 1+|R|^2+2|R|\cos(2kx+\delta)e^{-2x/l_\varphi} \qquad (2)$$

Eq. 2 provides an excellent fit to all of oscillatory amplitude vs. position data plotted at specific energies (Fig. 2b). Small deviations are also observed close to the position of the defect, although these are expected for our simple model and not crucial to the main point of our paper. More sophisticated calculations, which treat the defect in detail, should enable this point to be addressed clearly in the future.

Importantly, each fit provides a unique value of the wavevector and energy, and thus taken together, our data provide the desired energy band dispersion $E(k)$ vs. $k$ near $E_F$. Our experimental energy dispersion (Fig. 3a) agrees well with the linear dispersion predicted near $E_F$ for armchair SWNTs [4], $|E(k)|=(3^{1/2}/2)a\gamma_0|k-k_F|$, where a is the armchair lattice constant, $\gamma_0$ is the tight-binding overlap integral, and $k_F$ is the Fermi wavevector. A fit of our experimental energy dispersion with this expression yields $\gamma_0 = 2.51$ eV and $k_F = 8.48 \pm 0.05$ nm$^{-1}$. The experimental value of $\gamma_0$ is in good agreement with previous STM studies (2.5-2.7 eV) [3, 5-7] and first-principles calculations [4], the value of $k_F$ is also consistent with the prediction $k_F = 2\pi/3a = 8.52$ nm$^{-1}$ for an armchair SWNT [1]. The agreement of both parameters provides a strong consistency check in our work.

We have also compared the E($k$) vs. $k$ data obtained in our study with the band structure explicitly calculated for a (13,13) SWNT (Fig. 3b). The band structure was calculated using tight binding model described previously [3]. This calculation highlights the excellent agreement that our experimental data have with the linear $\pi^*$ band, but also shows that no data overlaps with the $\pi$ band. While at first glance surprising, this latter absence is completely consistent with the distinct symmetries of the $\pi$ and $\pi^*$ bands in armchair SWNTs. Specifically, the $\pi$ and $\pi^*$ bands of armchair SWNTs have opposite parity because they are constructed from rotationally invariant combinations of symmetric and antisymmetric sums of $p_z$ orbitals [4]. Therefore, if the defect observed in Fig. 1a has odd parity, electrons in the $\pi$ band (even parity) will not be scattered by the defect, whereas electrons in the $\pi^*$ band (odd parity) can be scattered when their energy matches one of the quasi-bound states of the defect. Several additional pieces of evidence support this interesting parity observation. First, the defect does not break the structural symmetry of the SWNT (both sides of the defect are consistent with a (13,13) armchair nanotube), and thus prevents mixing of $\pi$ and $\pi^*$ bands. Second, analysis of the position and energy dependent LDOS data obtained from the other side of the SWNT yields the same $\pi^*$ energy dispersion as plotted in Fig. 3 [18].

Lastly, we comment on the coherence lengths ($l_\varphi$) and scattering phase shifts ($\delta$) obtained from the analysis of data in Fig. 2b using Eq. 2. First, the average $l_\varphi$, 1.89 ± 0.27 nm, is substantially shorter than the values obtained from transport and optical measurements [1, 19-21] and implies a higher rate of phase-randomizing scattering. A source of the scattering may be $e$-$e$ interactions between electrons in the SWNT and Au(111) substrate. Because $k_F$(Au) is larger than $k_F$(SWNT), screening and $e$-$e$ interactions can occur on a much shorter length scale than in nanotubes supported on nonmetallic substrates, thus producing the short $l_\varphi$. Second, the average values of $\delta$ below and above E$_F$, –3.57 ± 0.13 and –2.06 ± 0.16, respectively, indicate a net repulsive interaction with the defect. These values also imply that the reflectivity, $|R|^2$, of the occupied defect states is close to unity, while $|R|^2$ is lower for the unoccupied defect states. We believe that the values of $l_\varphi$ and $\delta$ obtained in this work combined with more detailed theory should provide greater insight into the role of substrate-SWNT interaction and the contribution of individual defects to electrical resistance in SWNTs.

We have shown that resonant electron scattering by a defect provides a means to probe directly the 1D energy dispersion near E$_F$ in metallic SWNTs. Our results demonstrate clearly for the first time that the dispersion is linear in SWNTs as predicted by theory. In addition, these studies have highlighted the importance of parity—the specific defect structure—in the scattering in this molecular material. We believe that this observation opens up opportunities to investigate interesting questions about parity and symmetry breaking in SWNTs through investigations of scattering by different types of defects. Moreover, future investigations of resonant scattering provide a means for probing coherence, and thus should be useful for addressing $e$-$e$ interactions between nanotubes and the local environment—be it substrate or other nanotubes—and possibly $e$-$e$ interactions within a Luttinger liquid. The results from such fundamental studies will ultimately have important bearing on future applications of SWNTs.

We thank H. Park, J.F. Wang, L.J. Lauhon, K. Kim, W. Liang, L. Chen, M.S. Gudiksen for helpful discussions. C.M.L. acknowledges the support of this work by the NSF Division of Material Research.

**References & Notes**

size) system. In principle, systematic characterization of the wave periodicity for different eigenstates in finite size SWNTs could also provide a means for determining an effective energy dispersion, with the caveat that the details of the finite system may affect E(*k*).

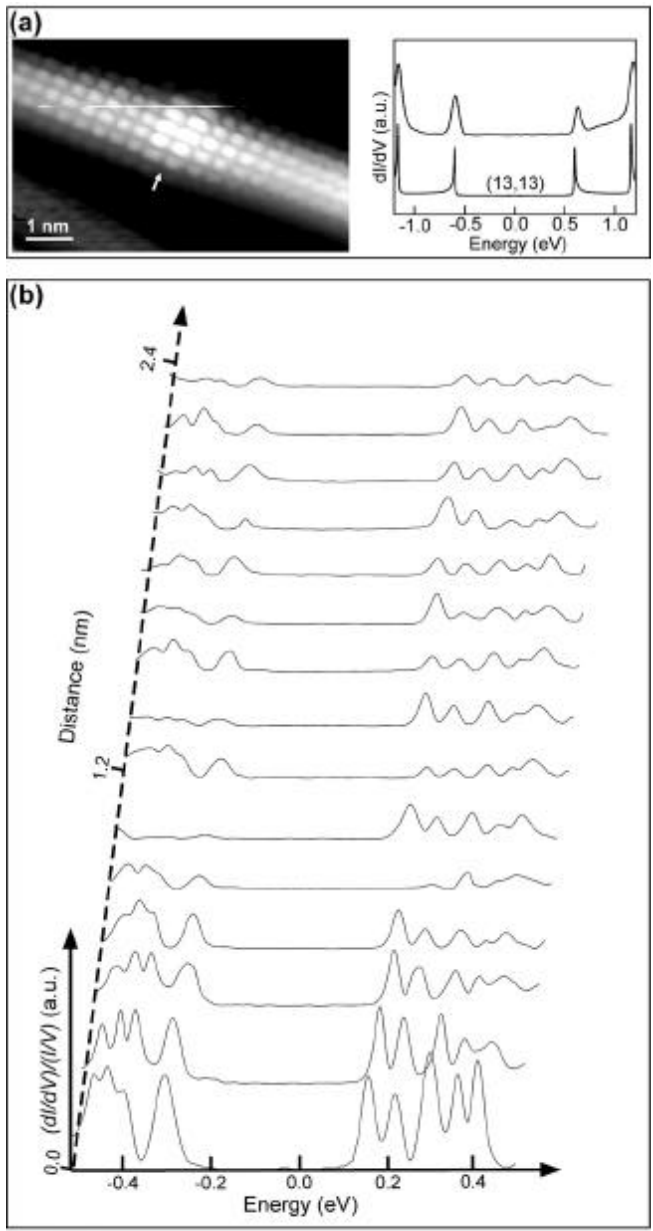

**FIG. 1.** (a) Atomically resolved STM image of an isolated armchair SWNT containing one defect; the defect position is highlighted with a white arrow. The image was recorded in the constant current mode with a bias voltage (*V*) of 0.55 V and a tunneling current (*I*) of 0.2 nA. (right) *(dI/dV)* vs. eV data recorded ca. 8 nm away from the defect and the LDOS for a (13,13) armchair SWNT calculated using a tight-binding model [3,7]. (b) Normalized tunneling conductance, *(dI/dV)/(I/V)* vs. eV recorded as a function of distance from the defect in (a). Curves start at the defect position and progress along the upper left portion of the nanotube. *(dI/dV)* data were recorded as the in-phase component of *I* using a lock-in amplifier with a 7.37-kHz, 2 mV peak-to-peak modulation of *V*.

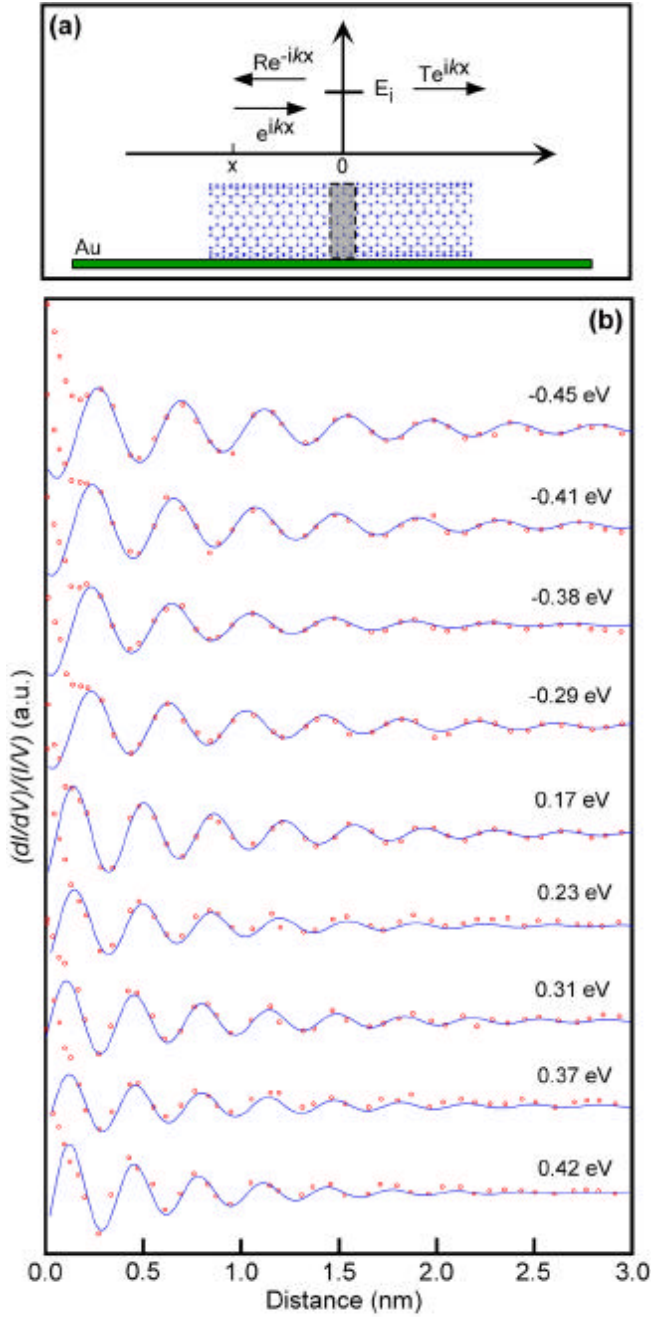

**FIG. 2.** (a) Schematic of electron scattering by a nanotube defect (shaded grey). When incident electrons with wavevector $k$ have the same energy as one of quasi-bound states, $E_i$, of the defect, the electron may be resonantly back scattered, with a reflectivity $|R|^2$, or transmitted with a coefficient $|T|^2$ ($|R|^2+|T|^2 =1$). (b) Spatial oscillation of *(dI/dV)/(I/V)* at all observed nine energies. The origin represents the defect position, which corresponds to the white arrow position in Fig. 1a. Red open circles: experimental data; Blue solid lines: theoretical fits of Eq. 2 to the data.

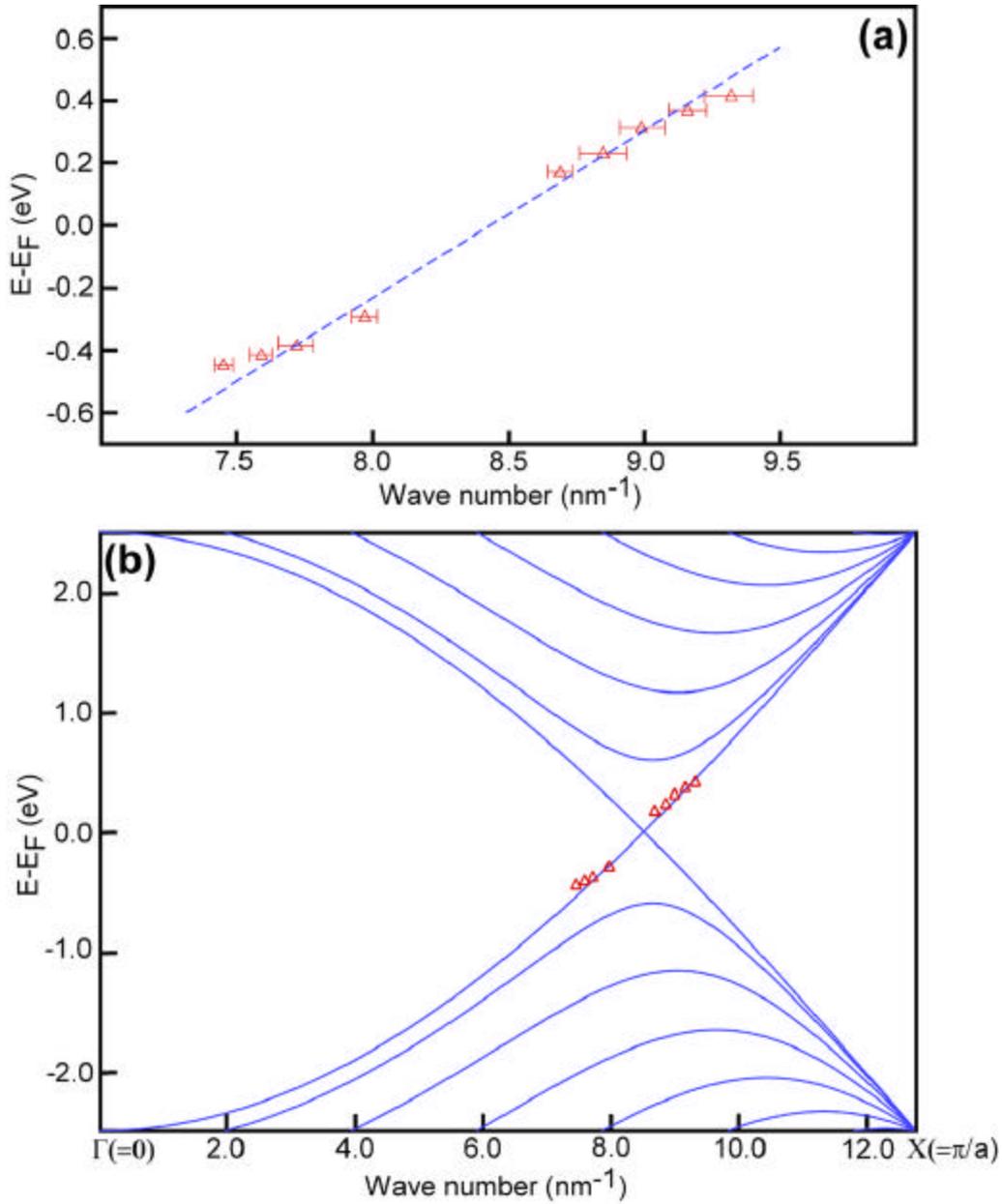

**FIG. 3.** (a) Energy dispersion near $E_F$ for the (13,13) armchair SWNT characterized in this study. The experimental data points (red triangles) fall on a straight dash line defined by $|E(k)|=(3^{1/2}/2)a\gamma_0|k-k_F|$, where the fitting parameters $\gamma_0$ and $k_F$ are described in the text. (b) Band structure of a (13,13) armchair SWNT plotted along the $\Gamma$-X line (blue solid lines) and experimental $E(k)$ vs. $k$ data points (red triangles).

8